# Science Platforms for Heliophysics Data Analysis

Monica G. Bobra (Stanford University), Will T. Barnes (NRC Postdoc residing at the Naval Research Laboratory), Thomas Y. Chen (Columbia University), Mark C. M. Cheung (Lockheed Martin Solar and Astrophysics Laboratory), Laura A. Hayes (NASA Goddard Space Flight Center), Jack Ireland (NASA Goddard Space Flight Center), Miho Janvier (Institut d'Astrophysique Spatiale), Michael S. F. Kirk (NASA Goddard Space Flight Center and ASTRA llc.), James P. Mason (University of Colorado Boulder), Stuart J. Mumford (The University of Sheffield and Aperio Software), Paul J. Wright (Stanford University)

## Introduction: What is the problem with data analysis in heliophysics today?

In 2020, we live in an era rich with data. NASA instruments designed to study the heliosphere take more data, at faster rates than ever before. For example, the Solar Dynamics Observatory (SDO) mission data set currently totals 19 petabytes. The heliophysics community also conducts coordinated campaigns, where many instruments observe the same target for the same time, that yield a wide variety of data—image data, spectral data, time series data, and *in situ* data on local particle and field conditions. Heliophysics is a naturally multi-messenger discipline, and many studies rely on using these varied data sets together.

However, scientists struggle to freely explore these data. Network speeds and limited storage make it difficult to obtain large data sets. Inadequate computing power makes it difficult to efficiently analyze large data sets. Despite the wide availability of accelerated and scalable computing resources (e.g. GPUs, cloud computing), most heliophysicists use their laptop or desktop computers, which are severely limited in both RAM and disk space, for their data analysis workflows. These hurdles prevent the community from maximizing scientific discovery and scientific return on investment.

In order to make new scientific breakthroughs with these data, we need to adopt a modern scientific workflow. In today's prevailing paradigm, heliophysicists download data to analyze on a local machine. This workflow limits the scope of scientific studies that rely on large data sets. Scientists only pursue studies feasible within the constraints of their local network speeds, storage capabilities, software tools, and computing power.

To support much larger studies, we recommend an alternate paradigm. In this workflow, heliophysicists conduct scientific studies on an external machine equipped with multiple mission data sets, ample computing power, and software tools. These openly accessible science platforms, now available in the fields of astrophysics (e.g. Bauer et al. 2019) and earth science (e.g. Robinson et al. 2019), allow scientists to rapidly analyze petabytes of data. Unburdened by network speeds, storage, software, and compute, heliophysicists can freely pursue compelling scientific studies that were previously practically impossible.

## Science platforms for other scientific communities

Other scientific communities already recognized the power of this approach. For example,

astronomers can analyze sky survey data from the Vera Rubin Observatory with a NSF-funded science platform that co-locates the survey data with high-performance computing at the National Center for Supercomputing Applications (Dubois-Felsmann et al. 2019). Practically, this means astronomers can open a terminal window or Jupyter Lab, log onto an external computing platform, write code, and run it on any of the survey data. Other examples, such as the NSF-funded science platform called the National Optical Astronomy Observatory Data Lab, exist as well (Fitzpatrick et al. 2014, Taghizadeh-Popp et al. 2020, Thomas et al. 2020).

The Pangeo data science platform (Odaka et al. 2020), funded by various institutions including the NSF, NASA, NCAR, Sloan Foundation, and UK Met Office, completely revolutionized the way many earth scientists think about data science workflows. Earth scientists use Pangeo to analyze petabytes from a variety of space-based observatories.

## The four necessary attributes of a science platform

While the technical design elements of each platform vary, each of them includes four essential attributes: data storage, computing power, software tools, and open access. The scientific data, version-controlled and stored in flexible databases, should adhere to a standard and adaptable format. Computational resources, co-located with the data, should allow parallel processing on CPUs and GPUs. Science platforms should also provide environments to easily install and use open-source, openly-developed, and version-controlled scientific software. Finally, the platform should provide openly accessible collaborative workspaces for the entire community.

## Science platforms for heliophysics

The heliophysics community will benefit immensely from a science platform. Heliophysics observatories produce extremely large data sets that are not used to their full potential. According to a survey of the solar physics community by the SunPy project (Bobra et al. 2020), 82% of solar physicists work with observational data.

Despite this, most heliophysicists lack access to computational facilities. In the same survey, the SunPy project found that 14% of the solar physics community uses local or regional clusters, 9% use GPUs, and 5% use the commercial cloud to do their research. Roughly a third of the community uses exclusively a laptop or desktop. This also means most of the community does not take advantage of massively parallel computing, even though it presents the biggest opportunity to accelerate computing performance (Robinson et al. 2020).

Open-source scientific software already provides powerful, easy-to-use tools that scalabely accelerate computing performance (e.g. Dask; Rocklin et al. 2015). Furthermore, open-source, version-controlled instrument calibration software, such as AIAPy (Barnes et al. 2020), co-located with open-access, version-controlled data can help the community create calibrated, reproducible, shareable data sets.

Today, science platforms do not exist in the heliophysics community. A solar science platform prototype (Barnes et al. 2019), which provides interactive access to SDO data and high-performance computing with the NASA Pleiades supercomputer, demonstrates how users can analyze large data sets quickly. Science platforms such as these will usher in a new era of scalable, interactive supercomputing for data analysis in solar and space physics.

## Recommendations

We recommend that NASA maintain and fund science platforms that enable interactive and scalable data analysis in order to maximize the scientific return of data collected from space-based instruments[1]. In support of this vision, we recommend a series of short-term goals to achieve within the next 10 years: (1) support coordinated and open development of scientific software that balances interactivity and scalability, (2) provide community education and training on high-performance and cloud computing for data analysis, and (3) establish a funding model where grant dollars can buy computing time from commercial cloud vendors.

We also recommend a series of long-term goals to achieve within the next 20 years: (1) establish dedicated infrastructure at NASA high-performance computing centers (e.g. ADAPT at Goddard, Pleiades at Ames) for interactive, scalable data analysis, and (2) ensure that the scientific potential of these data are maximized long past the lifetime of the mission by co-locating current and final mission data archives with this dedicated infrastructure. Together, these recommendations allow scientists to accelerate their research workflows, produce reproducible research, and maximize the scientific return of NASA data sets.

---

[1] This vision and these short- and long-term goals are consistent with Recommendation 3.2.4 in the 2020 National Academies report entitled *Progress Toward Implementation of the 2013 Decadal Survey for Solar and Space Physics: A Midterm Assessment*, that "NASA and NSF should maximize the scientific return from large and complex data sets by supporting (1) training opportunities on modern statistical and computational techniques; (2) science platforms to store, retrieve, and process data using common standards; (3) funding opportunities for interdisciplinary collaboration; and (4) the development of open-source software."